\documentclass[aps,prl,twocolumn,showpacs]{revtex4}

\usepackage{graphicx}
\usepackage{dcolumn}
\usepackage{bm}

\begin{document}

\title{Low-Temperature Heat Transport in the Low-Dimensional Quantum 
Magnet NiCl$_2$-4SC(NH$_2$)$_2$}

\author{X. F. Sun}
\email{xfsun@ustc.edu.cn}

\affiliation{Hefei National Laboratory for Physical Sciences at
Microscale, University of Science and Technology of China, Hefei, 
Anhui 230026, P. R. China}

\author{W. Tao}
\affiliation{Hefei National Laboratory for Physical Sciences at 
Microscale, University of Science and Technology of China, Hefei, 
Anhui 230026, P. R. China}

\author{X. M. Wang}
\affiliation{Hefei National Laboratory for Physical Sciences at
Microscale, University of Science and Technology of China, Hefei, 
Anhui 230026, P. R. China}

\author{C. Fan}
\affiliation{Hefei National Laboratory for Physical Sciences at 
Microscale, University of Science and Technology of China, Hefei, 
Anhui 230026, P. R. China}

\date{\today}

\begin{abstract}

We report a study of the low-temperature thermal conductivity of 
NiCl$_2$-4SC(NH$_2$)$_2$, which is a spin-1 chain system 
exhibiting the magnon Bose-Einstein condensation (BEC) in 
magnetic field. It is found that the low-$T$ thermal conductivity 
along the spin-chain direction shows strong anomalies at the 
lower and upper critical fields of the magnon BEC state. In this 
state, magnons act mainly as phonon scatterers at relatively high 
temperature, but change their role to heat carriers upon 
temperature approaching zero. The result demonstrates a large 
thermal conductivity in the magnon BEC state and points to a 
direct analog between the magnon BEC and the conventional one.

\end{abstract}


\maketitle

Bose-Einstein condensation (BEC) denotes the formation of a 
collective quantum ground state of identical particles obeying 
Bose statistics. This fascinating state of matter is well 
established for liquid $^4$He and ultra-cold alkali atoms. It 
turns out that a form of BEC can also be observed in quantum 
magnets \cite{Review, Review_MM, Affleck, Giamarchi, Rice}, in 
which the density of magnons (bosons) can be tuned by an external 
magnetic field (playing the role of chemical potential). 
Recently, this so-called magnon BEC state has been experimentally 
realized in a growing number of dimerized spin-1/2 systems, such 
as the three-dimensional system TlCuCl$_3$ \cite{Nikuni, Ruegg1}, 
the quasi-two-dimensional system BaCuSi$_2$O$_6$ \cite{Jaime, 
Sebastian}, and the spin-ladder compounds 
(CH$_3$)$_2$CHNH$_3$CuCl$_3$ and (C$_5$H$_{12}$N)$_2$CuBr$_4$ 
\cite{Garlea, Lorenz} etc. A common characteristic of these 
quantum magnets is the spin-gapped ground state at zero field. 
The external magnetic field can close the spin gap and lead to a 
long-range antiferromagnetic (AF) order when the Zeeman energy 
overcomes the gap between the singlet ground state and the 
excited triplet states. It is notable that some alternative 
theories have also been proposed to explain the field-induced AF 
ordered state in quantum magnets and challenges the the validity 
of magnon BEC scenario for these systems \cite{Bugrij, Kalita}.  
To firmly establish the BEC of magnons and to develop a deeper 
understanding of this novel state of matter, it would be 
desirable to look for obvious analogs in the basic physical 
properties between the magnon BEC and the conventional one. One 
of the outstanding properties of the superfluid $^4$He is the 
extremely large thermal conductivity ($\kappa$) \cite{Keesom, 
Wilks}, which is well understood using the two-fluid model. It is 
natural to ask whether the thermal conductivity (by magnons) in 
the magnon BEC state behaves similarly as that in the superfliud 
$^4$He. The first experimental exploration done by Kudo {\it et 
al.} \cite{Kudo} did reveal an enhancement of thermal 
conductivity at the magnon BEC transition of TlCuCl$_3$. However, 
without carefully studying the anisotropic heat transport, it is 
not clear whether the enhancement is caused by the weakening of 
phonon scattering or the appearing of magnetic heat carriers. 

\begin{figure}
\includegraphics[clip,width=7.5cm]{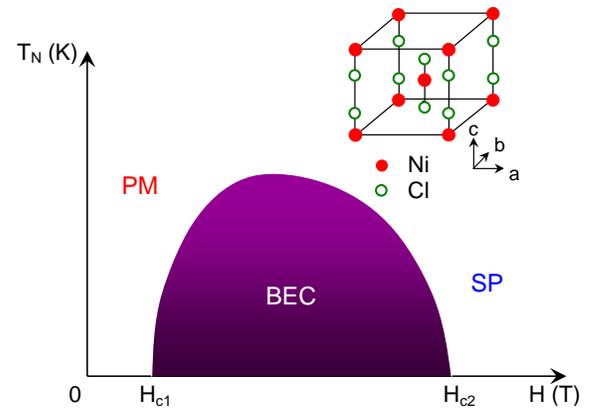}
\caption{(color online) Temperature-field phase diagram of DTN 
($H \parallel c$) obtained from magnetization, specific heat and 
magnetocaloric effect measurements \cite{Zapf1, Zvyagin1, Cox, 
Zapf2}. PM and SP represent the low-field quantum paramagnetic 
state and high-field spin-polarized state, respectively. Magnon 
BEC is a magnetic-field induced AF ordered state, with the lower 
and upper critical fields $H_{c1}$ ($\sim$ 2 T) and $H_{c2}$ 
($\sim$ 12 T). The maximum of field-dependent critical 
temperature is about 1.2 K, where $H_{c1}$ and $H_{c2}$ merge. 
Inset: Unit cell of tetragonal structure showing Ni and Cl atoms. 
The other atoms are omitted for clarity.}
\end{figure}

The organic compound NiCl$_2$-4SC(NH$_2$)$_2$ [dichloro-tetrakis 
thiourea-nickel (II), abbreviated as DTN] is the only quantum 
spin-1 system, rather than spin-1/2 dimers, to exhibit the BEC of 
spin degrees of freedom \cite{Zapf1, Zvyagin1, Cox, Zapf2, Yin, 
Zvyagin2}. It has a tetragonal crystal structure (space group I4) 
\cite{structure}, which satisfies the axial spin symmetry 
requirement for a BEC. The Ni spins are strongly coupled along 
the tetragonal $c$ axis (Fig. 1), making DTN a system of weakly 
interacting spin-1 chains with single-ion anisotropy larger than 
the intra-chain exchange coupling. The anisotropy, intra-chain 
and inter-chain exchange parameters of Ni spins were determined 
to be $D$ = 8.9 K, $J_c$ = 2.2 K and $J_{a,b}$ = 0.18 K 
\cite{Zapf1, Zvyagin1, Cox, Zapf2}, respectively. It was found 
that the Ni spin triplet is split into a $S_z$ = 0 ground state 
and $S_z = \pm$1 excited states with an anisotropy gap of $D$, 
which precludes any magnetic order at zero field. When a magnetic 
field is applied along the $c$ axis, the Zeeman effect lowers the 
$S_z$ = 1 level until it becomes degenerate with $S_z$ = 0 ground 
state at $H_{c1}$, which is essentially the same as that occurs 
in the spin-1/2 dimerized systems. Between $H_{c1}$ and $H_{c2}$ 
and below the maximum $T_c(H)\sim$ 1.2 K (Fig. 1), the magnetic 
field induces an AF order or a magnon BEC. For magnetic field 
perpendicular to the $c$ axis, however, the $S_z$ = 0 ground 
state mixes with a linear combination of the $S_z = \pm$1 excited 
states and there is no level crossing with increasing field and 
therefore no magnetic order \cite{Zapf1}. DTN was found to be an 
ideal system for studying the magnon BEC in the sense that its 
lower and upper critical fields are not high, about $H_{c1} \sim$ 
2 T and $H_{c2} \sim$ 12 T, which are easily achievable by the 
common laboratory magnets. In this Letter, we show that the 
low-temperature and high-magnetic-field thermal conductivity of 
DTN single crystals indeed demonstrates a large thermal 
conductivity in the magnon BEC state.

High-quality NiCl$_2$-4SC(NH$_2$)$_2$ single crystals are grown 
from aqueous solution of thiourea and nickel chloride 
\cite{structure}. The typical size of single crystals is (0.5--2) 
$\times$ (0.5--2) $\times$ (3--4) mm$^3$. X-ray diffraction 
indicates that the parallelepiped crystals are grown along the 
$c$ axis (the maximum dimension) while the four side surfaces are 
the (110) crystallographic plane. So it is easy to prepare 
samples for the thermal conductivity measurements either along 
($\kappa_c$) or perpendicular to ($\kappa_{ab}$) the direction of 
spin chains. The thermal conductivity is measured using a 
conventional steady-state technique and two different processes: 
(i) using a ``one heater, two thermometers" technique in a $^3$He 
refrigerator and a 14 T magnet for taking data at temperature 
regime of 0.3--8 K; (ii) using a Chromel-Constantan thermocouple 
in a $^4$He cryostat for taking zero-field data above 4 K 
\cite{Sun}. It is worthy of emphasizing that a careful 
pre-calibration of resistor sensors is indispensable for the 
precise thermal conductivity measurements in high magnetic fields 
and at low temperatures.

\begin{figure}
\includegraphics[clip,width=8.5cm]{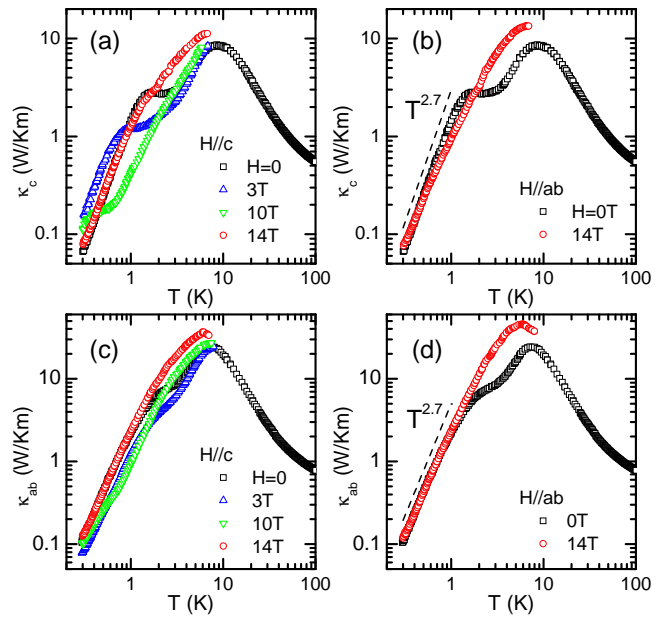}
\caption{(color online) Temperature dependences of thermal 
conductivities of DTN single crystals for $\kappa_c$ and 
$\kappa_{ab}$ in both the zero field and several different 
magnetic fields up to 14 T, which applied along the $c$ axis or 
the $ab$ plane. The dashed lines in panels (b) and (d) indicate 
that the zero-field thermal conductivities at subKelvin 
temperatures show a $T^{2.7}$ temperature dependence.}
\end{figure}

Figure 2 shows the temperature dependences of $\kappa_c$ and 
$\kappa_{ab}$ in zero field and several magnetic fields up to 14 
T. Like in usual insulating crystals \cite{Berman}, there is a 
clear phonon peak at 8--9 K in both $\kappa_c$ and $\kappa_{ab}$, 
for which the peak magnitude is 8.5 and 24 W/Km, respectively. It 
can be seen that the heat conductivity of this organic crystal is 
rather large compared to common organic materials; actually, the 
phonon peak in DTN is comparable or even larger than that in many 
inorganic crystals, like transition-metal oxides. One remarkable 
behavior of $\kappa(T)$ in zero field is a ``shoulder"-like 
feature at $\sim$ 2K; in addition, the ``shoulder" moves to lower 
temperature upon increasing the magnetic field. This kind of 
temperature dependence in $\kappa(T)$ usually indicates a 
resonant phonon scattering \cite{Berman} by some lattice defects, 
magnetic impurities, or magnon excitations, etc. Apparently, the 
sensitivity of the ``shoulder" to the applied field suggests the 
magnetic origin of this resonant scattering. Another important 
feature of $\kappa(T)$ is that at subKelvin temperature regime, 
the thermal conductivity show a $T^{2.7}$ dependence, which is 
close to the $T^3$ law of the standard temperature dependence of 
the phonon thermal conductivity in the boundary scattering limit 
\cite{Berman}. One possible reason of the slight deviations from 
$T^3$ law is due to the phonon specular reflections at the sample 
surface \cite{Pohl, Sutherland}.  

\begin{figure}
\includegraphics[clip,width=8.5cm]{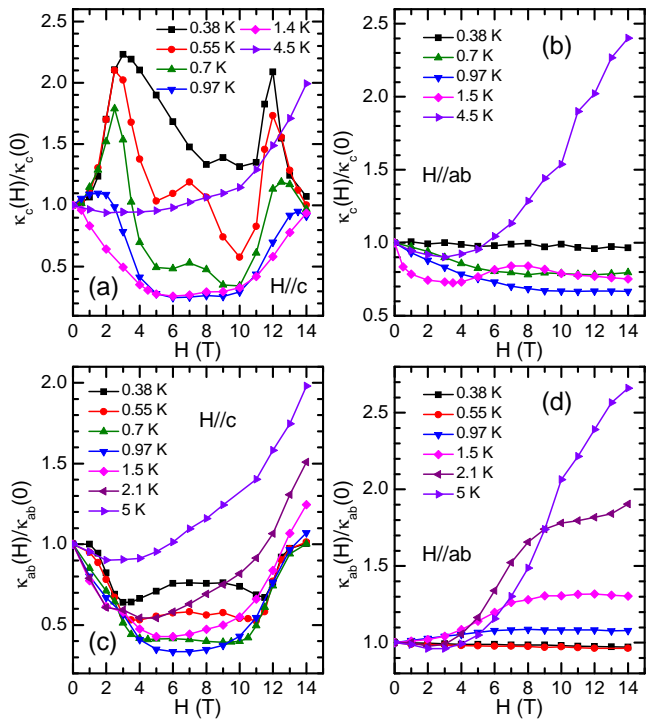}
\caption{(color online) Magnetic-field dependences of thermal 
conductivities of DTN single crystals at low temperatures.}
\end{figure}

Detailed magnetic-field dependence of the low-temperature thermal 
conductivity is a key to understanding the mechanism of heat 
transport in the low-dimensional spin systems \cite{Ando}. Figure 
3 shows the $\kappa(H)$ isotherms for both $\kappa_c$ and 
$\kappa_{ab}$; in each case the magnetic field is applied both 
along the $c$ axis and along the $ab$ plane. Although the 
$\kappa(H)$ behaviors in general are rather complicated in these 
four measurement configurations, one can easily notice the most 
striking result shown in Fig. 3(a), that is, at very low 
temperatures $\kappa_c(H)$ display two peak-like anomalies across 
$\sim$ 2.5 T and 12 T ($\parallel c$), which are very close to 
the reported critical fields $H_{c1}$ and $H_{c2}$ \cite{Zapf1, 
Zvyagin1, Cox}, and the anomalies are getting enhanced upon 
temperature approaching zero. Furthermore, $\kappa_{ab}(H)$ also 
show steep changes across these two characteristic fields for $H 
\parallel c$. The interesting point is that, both $\kappa_c(H)$ and 
$\kappa_{ab}(H)$ do not show any drastic change at $\sim$ 2.5 T 
and 12 T for $H \perp c$. Since DTN does not exhibit any 
field-induced magnetic ordering or magnon BEC state when the 
magnetic field is perpendicular to the $c$ axis \cite{Zapf1}, 
these data clearly indicate that the strong peak-like anomalies 
in Fig. 3(a) are related to the quantum phase transitions at 
$H_{c1}$ and $H_{c2}$. 

\begin{figure}
\includegraphics[clip,width=4.5cm]{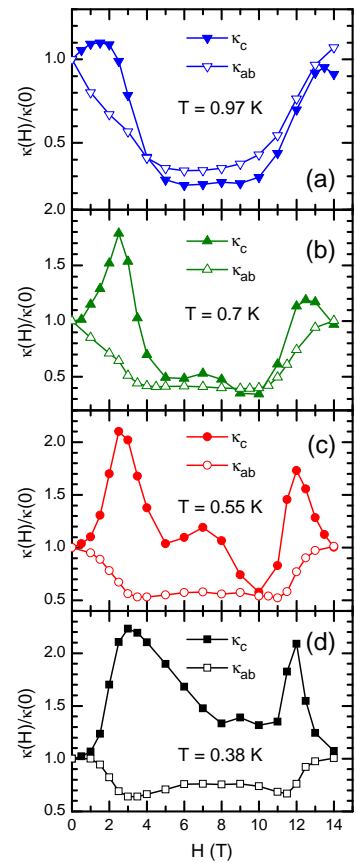}
\caption{(color online) Comparison of $\kappa_c(H)$ and 
$\kappa_{ab}(H)$ isotherms for $H \parallel c$ at subKelvin 
temperatures. (a) $T$ = 0.97 K. (b) $T$ = 0.7 K. (c) $T$ = 0.55 
K. (d) $T$ = 0.38 K.}
\end{figure}

It is interesting to compare $\kappa_c(H)$ and $\kappa_{ab}(H)$ 
behaviors in the $c$-axis field, in which the magnon BEC can 
occur. Above 1.4 K, $\kappa_c$ and $\kappa_{ab}$ have essentially 
similar magnetic-field dependences, as shown in Figs. 3(a) and 
3(c), while the difference between them shows up and becomes 
larger upon lowering temperature. For clarity, Fig. 4 shows a 
direct comparison of $\kappa_c(H)$ and $\kappa_{ab}(H)$ at 
subKelvin temperatures. At 0.97 K, both $\kappa_c(H)$ and 
$\kappa_{ab}(H)$ show a ``U"-shaped curve: a steep decrease 
across 2.5 T, a strong suppression but weak field dependence in 
the intermediate field regime, and a steep recovery of 
conductivity across 12 T. There is also a small difference 
between these two curves, that is, a small and broad peak below 2 
T shows up in $\kappa_c(H)$ isotherm. With lowering temperature, 
the behavior of $\kappa_{ab}(H)$ does not change much, except 
that the suppression of the thermal conductivity in the 
intermediate field regime is gradually getting weaker and two 
shallow ``dips" appear at $\sim$ 3 T and 11.5 T. In the meantime, 
the behavior of $\kappa_c(H)$ changes much more drastically. 
First, the large peak-like anomalies at $\sim$ 2.5 T and 12 T 
show up below 0.7 K and become more significant upon $T \to$ 0 K. 
Second, the suppression of thermal conductivity in the 
intermediate field regime is getting weaker rather rapidly with 
lowering temperature and finally evolves to an enhancement 
(compared to the zero-field conductivity) at 0.38 K, which 
strongly suggests that there are two competing impacts on thermal 
conductivity induced by the magnetic field. Apparently, the main 
difference between $\kappa_c$ and $\kappa_{ab}$ can only come 
from anisotropic magnetic contributions to heat transport, acting 
as either heat carriers or phonon scatterers. Because of the 
strong anisotropy of the magnon dispersion \cite{Zapf1}, it is a 
natural conclusion that the strong suppression of 
$\kappa_{ab}(H)$ at $H_{c1} < H < H_{c2}$ is mainly due to the 
phonon scattering by magnons; on the other hand, although the 
magnon scattering can also weaken the phonon thermal conductivity 
along the $c$ axis, the magnons (with stronger dispersion in this 
direction) can act as heat carriers and make an additional 
contribution to the heat transport. Furthermore, between the two 
competing roles of magnons in affecting $\kappa_c$, i.e., 
scattering phonons or carrying heat, the latter one is apparently 
dominant at low temperatures. Note that because of the two 
competing effects of magnons on the heat transport, the ability 
of magnons to carry heat must be much larger than what the raw 
$\kappa_c(H)$ data ($H \parallel c$) demonstrate. To our 
knowledge, there has been no such clear evidence showing a large 
thermal conductivity in the magnon BEC state. 

Besides the above clear information demonstrated by the 
anisotropic heat transport behaviors, one may notice that the 
details of the $\kappa(H)$ data are actually rather complicated 
and some considerable further investigations are needed. In 
principal, the competing roles of magnons acting as phonon 
scatterers and as heat carriers may lead to complicated field 
dependence of $\kappa$ ($H \parallel c$), including the peak-like 
anomalies of $\kappa_c(H)$ at $H_{c1}$ and $H_{c2}$ and the local 
maximum between two peaks. Besides, the peak-like anomalies can 
be closely related to the maximized magnon population at $H_{c1}$ 
and $H_{c2}$, since the dispersion becomes quadratic at the 
critical fields while it is linear for $H_{c1} < H < H_{c2}$ 
\cite{Matsumoto}; another contribution to the non-monotonic field 
dependence of $\kappa_c(H)$ between $H_{c1}$ and $H_{c2}$ may be 
coming from some upper magnon branches having rather small gap at 
the intermediate field \cite{Zvyagin2}. On the other hand, the 
$\kappa_c(H)$ and $\kappa_{ab}(H)$ for $H \perp c$ are also 
rather complicated. In general, the field dependences of $\kappa$ 
at very low temperature are rather weak, consistent with the fact 
that for $H \perp c$ there is no field-induced transition and the 
magnetic excitations are always gapped \cite{Zapf1, Zapf2}. The 
most drastic field dependence is the strong increase of 
$\kappa_{ab}$ and $\kappa_c$ at 4--5 K, which is actually very 
similar to the behaviors in $H \parallel c$. This is probably 
because in such high temperature region, where the temperature 
scale is comparable to the spin anisotropy gap, there are strong 
phonon scattering caused by magnetic excitations (remember the 
resonant scattering feature of $\kappa(T)$ in zero field shown in 
Fig. 2), which can be weakened when applied field increases the 
energy of magnetic excitations \cite{Zvyagin1, Cox}. 

It is intriguing to point out that the experimental phenomenon 
cannot be simply explained as the strong magnetic heat transport 
found in some low-dimensional spin systems \cite{Hess}. Actually, 
it was originally predicted that the spin transport is diffusive 
and finite in the spin-1 chain material since it is not an 
integrable system \cite{Haldane}, while the experimental results 
were quite controversial, both low and high magnon thermal 
conductivities have been observed in different compounds 
\cite{Sologubenko1, Kordonis, Sologubenko2}. It is notable that 
in the magnon BEC state, the spin system is no longer one 
dimensional; instead, it is a three-dimensional ordered state 
\cite{Zapf1, Zvyagin1, Cox}. Because of the crossover of the 
dimensionality and the character of magnetic quasiparticles at the
BEC transition \cite{Chiatti}, it is natural to expect a 
different mechanism of magnon heat transport from that of the 
low-dimensional systems when the field-induced long-range 
magnetic order is established. 

To summarize the main finding of this work, the magnon heat 
transport is found to be large in the magnon BEC state of DTN 
single crystals, pointing to a direct analogy between the magnon 
BEC and the conventional one. Since the BEC condensate does not 
carry entropy, the large low-$T$ magnon heat transport can only 
be related to the uncondensed part of magnons. An elaborate 
theory, probably based on the two-fluid model established for 
superfluid $^4$He \cite{Wilks}, is called for quantitatively 
describing the heat transport in this novel state of quantum 
magnet. 

We thank Y. Ando, A. N. Lavrov and J. Takeya for helpful 
discussions. This work was supported by the Chinese Academy of 
Sciences, the National Natural Science Foundation of China (Grant 
Nos. 10774137 and 50721061) and the National Basic Research 
Program of China (Grant Nos. 2009CB929502 and 2006CB922005).

\end{document}